\RequirePackage{fix-cm}
\documentclass[twocolumn]{svjour3}          

\usepackage{graphicx}
\usepackage{amsmath}
\usepackage{enumerate}
\usepackage{epstopdf}
\usepackage{multirow}
\usepackage{array}
\usepackage{float}
\usepackage{subfigure}

\begin{document}

\title{Phone2Cloud: Exploiting Computation Offloading for Energy Saving on Smartphones in Mobile Cloud Computing}

\author{Feng Xia         \and
        Fangwei Ding     \and
        Jie Li \and
        Xiangjie Kong  \and
        Laurence T. Yang \and
        Jianhua Ma
}
\institute{Feng Xia, Fangwei Ding, Jie Li, Xiangjie Kong \at
              School of Software, Dalian University of Technology, Dalian 116620, China. \\
              \email{f.xia@ieee.org}           
           \and
           Laurence T. Yang \at
           Department of Computer Science, St. Francis Xavier University, Canada. \\
           \email{ltyang@stfx.ca}
           \and
           Jianhua Ma \at
           Faculty of Computer and Information Sciences, Hosei University, Japan. \\
           \email{jianhua@hosei.ac.jp}
           \and
           *Corresponding author: Xiangjie Kong (E-mail: xjkong@ieee.org).
}

\date{Received: 15 January 2013 / Accepted: date}

\maketitle

\begin{abstract}
With prosperity of applications on smartphones, energy saving for smartphones has drawn increasing attention. In this paper we devise Phone2Cloud, a computation offloading-based system for energy saving on smartphones in the context of mobile cloud computing. Phone2Cloud offloads computation of an application running on smartphones to the cloud. The objective is to improve energy efficiency of smartphones and at the same time, enhance the application's performance through reducing its execution time. In this way, the user's experience can be improved. We implement the prototype of Phone2Cloud on Android and Hadoop environment. Two sets of experiments, including application experiments and scenario experiments, are conducted to evaluate the system. The experimental results show that Phone2Cloud can effectively save energy for smartphones and reduce the application's execution time.
\end{abstract}
\section{Introduction}
\label{intro}
Smartphones have become increasingly popular in our daily life. They offer users more powerful processors, larger memory, multi-network interfaces and a wide variety of operating systems such as Apple iOS, Android and Windows Phone. Not only have the processor speed and memory size of smartphone increased significantly, but also the resolution of screen and the quality of the available sensors \cite{1}. The smartphone's processor speed has grown enormously in recent years as well as memory size, which have arrived at the same level as notebooks of five years ago. Furthermore, cellular networking technology grows from GSM networks to current 4G networks, which significantly increases bandwidth of wireless networks.

As processors are getting faster, memory is getting larger, screens are getting sharper and devices are equipped with multiple sensors \cite{2}, a large spectrum of novel and innovative applications have appeared. They are ranging from mobile games, to multimedia applications, social networking services and more \cite{3}. Currently, users can easily get applications from market places, like Apple App Store and Google Play. Additionally, users are more likely to run resource-demanding applications, such as rich media applications using multiple inputs like cameras and sensors \cite{4}. These applications imply a heavy workload on processors, wireless network interfaces and display, which causes a significant energy cost \cite{5}.

To sum up, smartphones provide multi-core processors, sharper screens, larger memory, multiple sensors and radios as well as enormous applications. These together put a heavy burden on battery's energy consumption \cite{1}. In the meantime, advances in battery technology and energy saving solutions have not kept pace with rapidly growing energy demands \cite{6}. Furthermore, smartphones are getting thinner and smaller, which implies smaller batteries and less battery capacities. Therefore, the energy consumption has always been primary bottleneck for smartphones.

Many researchers and engineers have made great efforts \cite{7,8,9,10,11,12,13} on saving energy to extend lifetime of batteries. One popular approach of energy saving for mobile devices is computation offloading: applications take advantage of resource-rich infrastructures by migrating computation to these infrastructures \cite{2}. Recently, researchers have recognized offloading computation via networks to cloud can help reduce power consumption of smartphones \cite{12,13,14,15,16,17}.

In this paper, we develop a computation offloading-based system for energy saving on smartphones, called Phone2Cloud. It offloads all or part of an application running on smartphones to cloud to achieve energy conservation, reduce the application's execution time and improve user's experience, i.e. meet user's delay-tolerance threshold as described in \cite{18}.

Here we make the following contributions. First, we develop the Phone2Cloud system for energy saving on smartphones. Second, we propose and implement a modified offloading-decision making algorithm based on \cite{14,19,20} in our system, and a new element - user's delay-tolerance threshold is involved in the offloading-decision making algorithm. Finally, we conduct two types of experiments on our system, and the results demonstrate the superiority of our system.

The rest of the paper is organized as follows. In next section, we review related works on reducing power consumption for smartphones, especially on those taking advantage of computation offloading. Section 3 presents the architecture of the Phone2Cloud system. In Section 4, the major algorithms employed in Phone2Cloud are described. Evaluation and analysis of Phone2Cloud are conducted in Section 5. We conclude the paper and discuss extensions to our system in Section 6.
\section{Related work}
\label{sec:2}
In this section, we briefly review four basic approaches for energy conservation and extending battery lifetime in smartphones.

\textbf{Smart battery models and energy cost models.} To efficiently and effectively use a battery, it is important to treat the battery as a measurable resource whose attributes are available to the operating system and applications. For this purpose, smart battery models and energy cost models have been presented to model battery's attributes. A number of battery models \cite{21} such as ideal model, stochastic model, diffusion model and so forth have been proposed. As to energy cost models, for instance, Kim \textit{et al.} \cite{22} present a low-level energy cost model for fast estimation of software energy consumption in off-the-shelf processor. Mahmud \textit{et al.} \cite{23} have proposed a high-level energy cost model for predicting energy consumption in the wireless network access portion of a handheld device equipped with multiple radio interfaces. Moreover, a system-level energy model is proposed by Palit \textit{et al.} in \cite{24}.

\textbf{Avoiding energy waste.} In this line of research, the whole system or individual component is put into sleep state to save energy. Brakmo \textit{et al.} \cite{25} present an energy reduction technique for handheld devices, called $\mu$Sleep. It tries to put processor in sleep mode for short periods to save energy without affecting the user's experience. However, it is hard to precisely predict when to enter sleep state. Shih \textit{et al.} \cite{26} propose an energy saving strategy, called wake on wireless, to reduce the phone's idle power. They power off the phone and its radio interfaces when the phone is not being used, and the phone is powered only when there is on-going traffic. Nevertheless users will completely lose network connection when the phone is not used.

\textbf{Communication related energy saving.} Many works \cite{27,28,29} on reducing the energy of network communications have been done. Zhang \textit{et al.} \cite{27} present and evaluate a system-level power management method for a mobile device to dynamically shut down its Wi-Fi interface. However, the idle state of Wi-Fi radio interface is hard to be predicted. Since Bluetooth causes much less energy than Wi-Fi, Pering \textit{et al.} \cite{28} develop a system called Coolspots that automatically switch between Wi-Fi and Bluetooth to increase battery lifetime. However, the system needs to modify the infrastructure. Blue-Fi \cite{29} uses a combination of Bluetooth contact-patterns and cell-tower information to predict the availability of the Wi-Fi connectivity, thus it avoids the long periods in idle state of Wi-Fi interface and significantly reduces the number of scans for Wi-Fi discovery. However, it cannot work very well outside due to short range of Bluetooth.

\textbf{Computation offloading-based energy saving.} The main idea in computation offloading is to migrate computation-intensive tasks from mobile device to a server or cloud via network in order to save energy on the mobile device. Quite a lot of works have been done on computation offloading, for example, \cite{1,14,15,16,17,19,30,31,32,33}. Most of them deal with offloading computation from mobile device to a desktop computer or server on the network. For example, Gu \textit{et al.} \cite{30} develop an adaptive offloading system. It dynamically partitions an application and efficiently offloads part of the application to be executed on a nearby server. However, they need to modify JVM to support transparent migration of objects between mobile device and server. In \cite{1}, Cuervo \textit{et al.} present MAUI, a system enabling fine-grained energy-aware offloading of mobile code to the infrastructure. Although MAUI's energy savings and performance are impressive, it still needs to partition applications and incurs extra overhead. Some others focus on offloading computation to cloud via network. For instance, Kumar and Lu \cite{14} conduct a qualitative analysis on whether cloud computing can extend battery lifetime for users. They believe cloud computing can potentially save energy for mobile users. Kemp \textit{et al.} \cite{15} study how smartphones can benefit from the resources available in cloud. They are building a framework for applications to be offloaded to cloud. However, there is not any quantitative analysis in these papers. In \cite{16}, Miettinen and Nurminen provide a quantitative analysis of the critical factors affecting the energy consumption of mobile clients in cloud computing. Aggarwal \textit{et al.} \cite{17} also conduct a quantitative analysis on mobile communication using cloud support. However, extensive quantitative analysis is still missing, and users' requirements, such as delay-tolerance threshold, are yet to be considered.

In contrast to the above state-of-the-art works, we focus on developing a computation offloading-based system and conducting a fully quantitative analysis on energy saving of the system. In addition, users' delay-tolerance threshold will be considered in our work.
\section{System architecture}
\label{sec:3}
An illustration of the architecture of Phone2Cloud is provided in Fig. 1. Phone2Cloud consists of seven key components, including a bandwidth monitor, a resource monitor, an execution time predictor, an offloading decision engine, a local execution manager, a remote execution manager and an offloading proxy that ties the offloading decision engine and remote execution manager together. Among these components, the offloading decision engine is the core and the area of extensibility in the architecture. Offloading decision mechanisms can be easily added or removed from the framework. We will describe these components in the following sub-sections.
\begin{figure}
\centering
\includegraphics [width=2.5in]{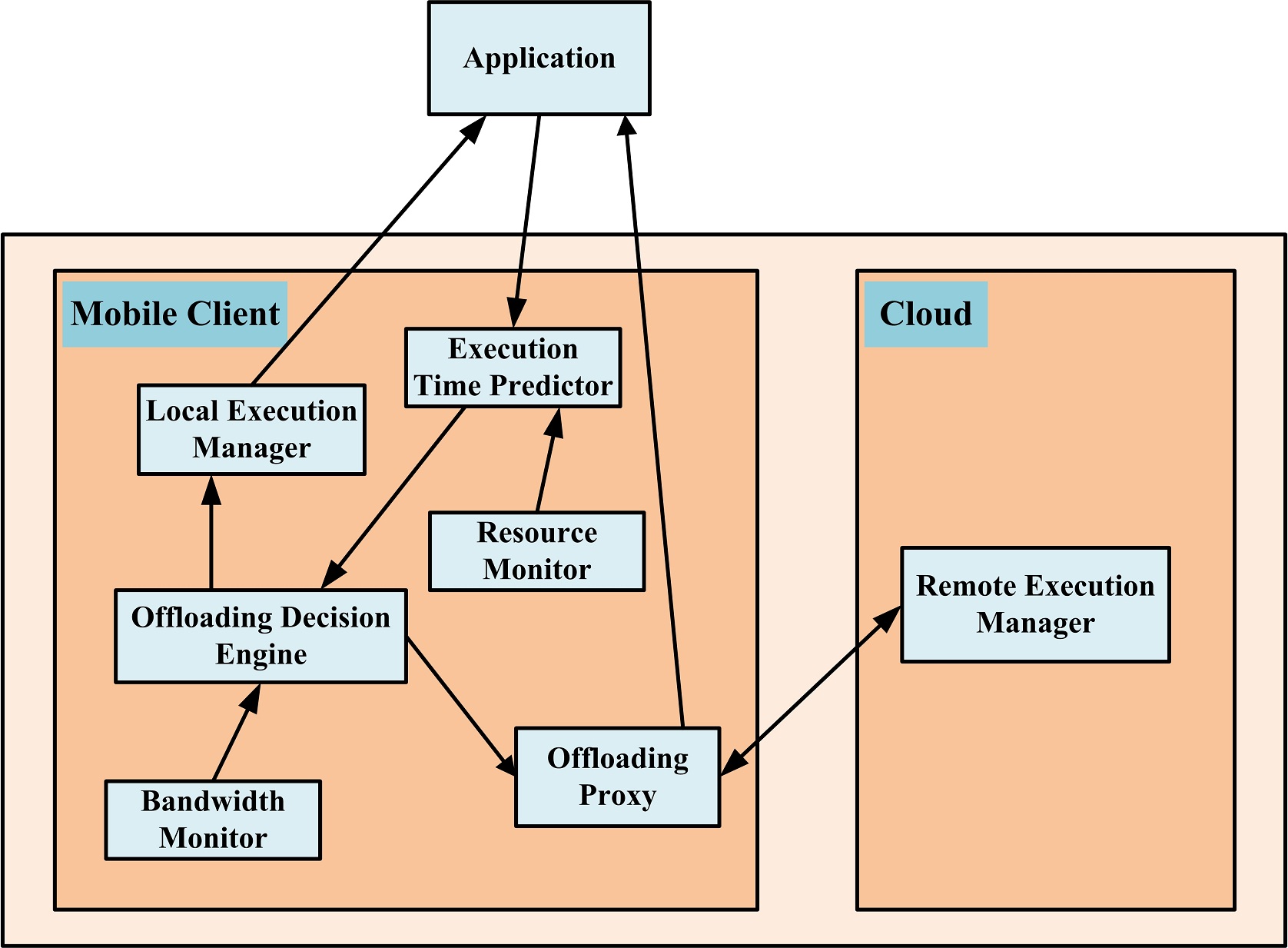}
\caption{Architecture of Phone2Cloud}
\label{fig:1}
\end{figure}
\subsection{Bandwidth monitor and resource monitor}
\label{sec:3-1}
In order to support making an offloading decision, we need to monitor current bandwidth usage of the network and status of the smartphone, such as CPU workload. Therefore we employ the bandwidth monitor and the resource monitor. The former periodically monitors bandwidth of current network to predict average bandwidth when making an offloading decision. The latter is used to monitor the status of smartphone. For sake of simplicity, we leverage it to monitor CPU workload of smartphone and predict the average CPU workload when offloading computation from smartphone to cloud. These two monitors serve the offloading decision engine and the execution time predictor separately, which will be discussed in detail below.
\subsection{Execution time predictor}
\label{sec:3-2}
The execution time predictor is one of the key parts in Phone2Cloud. It is used to predict average execution time of an entire application on smartphone. Many works \cite{34,35,36,37} have been conducted on execution time prediction. In order to simplify the complexity of Phone2Cloud, we use a simple history-based method to predict execution time of an application running on smartphone. Given an application, firstly, it gets the input size of the application and triggers the resource monitor to get predicted average CPU workload; secondly, it leverages the input size and the average CPU workload to search the log \textit{L} to find two nearest points in distance, and then returns the mean of the two points' execution times as the average execution time of the application. Here, the log \textit{L} is the execution history log and can be easily got by using a simple tool which repeatedly runs applications on smartphone. Besides, it is full of data in fixed form (\textit{Application, Input size, Average CPU workload, Execution time}), and we see entries of a specified application in log \textit{L} as points in a three-dimension space of input size, average CPU workload and execution time.
\subsection{Offloading decision engine}
\label{sec:3-3}
In this component, Phone2Cloud decides whether to offload the whole or part of an application to cloud, namely, whether to offload computation of the application from smartphone to cloud. Given an application, it calls the bandwidth monitor to get average bandwidth of current network, triggers the execution time predictor to get average execution time of the application, and then uses the offloading-decision making algorithm described in Section 4.3 to make an offloading decision. When it decides to run the application locally, it calls local execution manager to execute the application. Otherwise, it invokes offloading proxy to handle offloading computation to cloud.
\subsection{Offloading proxy}
\label{sec:3-4}
The offloading proxy sends required input data to the remote execution manager, receives the results returned by the remote execution manager, and delivers the results to the application. However, due to the fact that the application on smartphone cannot be directly run on cloud, we need to manually modify the application to make it possible not only to be run on cloud, but also to receive results from the offloading proxy. As we can see, Phone2Cloud is a semi-automatic offloading system.
\subsection{Local execution manager and remote execution manager}
\label{sec:3-5}
The local and remote execution managers are mainly used to manage execution of the application in Phone2-Cloud. The local execution manager is designed to execute the application on smartphones, simply invoking smartphone's operating system, such as Android and iOS, to run the application and logs the execution information into log \textit{L}. When the remote execution manger gets required input data from the offloading proxy, it executes offloading computation on cloud, and returns results to the offloading proxy.
\section{Algorithms}
\label{sec:4}
In this section, we will describe the key methods used in Phone2Cloud, including CPU workload prediction in the resource monitor, bandwidth prediction in the bandwidth monitor, and the offloading-decision making algorithm.
\subsection{CPU workload prediction}
\label{sec:4-1}
As discussed above, we propose a naive history-based method to predict average execution time of an application on smartphone. It leverages average CPU workload got from the resource monitor and input size of the application to predict execution time using the history log \textit{L}.

As to CPU workload prediction, the resource monitor uses the basic exponential moving average algorithm (EMA for short) \cite{38} to predict the average CPU workload of smartphone. It records the CPU workload $c_{t}$ in database in form (\textit{Timestamp, CPU workload}) periodically. Given current time period \textit{t}, the EMA value for CPU workload is calculated recursively by (1), (2) and (3), where $C_{t}$ is the value of the EMA at any time period \textit{t}, coefficient $\alpha$ represents the degree of weighting decrease and $N$ is the number of time periods. We simply use the EMA value $C_{t}$ as the average CPU workload.
\begin{equation}
C_{1}=c_{1}
\end{equation}
\begin{equation}
C_{t}=\alpha\cdot c_{t}+(1-\alpha)\cdot C_{t-1}
\end{equation}
\begin{equation}
\alpha=2/(N+1)
\end{equation}
\subsection{Bandwidth prediction}
\label{sec:4-2}
To predict average bandwidth, EMA is also used in the bandwidth monitor when making an offloading decision. Moreover, it uses (4), (5) and (6) to recursively calculate EMA value $B_{t}$ for bandwidth at current time period \textit{t}, where $b_{t}$ is the bandwidth recorded by bandwidth monitor periodically, coefficient $\beta$ has the same meaning with $\alpha$, and $N$ is also the number of time periods.
\begin{equation}
B_{1}=b_{1}
\end{equation}
\begin{equation}
B_{t}=\beta\cdot b_{t}+(1-\beta)\cdot B_{t-1}
\end{equation}
\begin{equation}
\beta=2/(N+1)
\end{equation}
\subsection{Offloading-decision making algorithm}
\label{sec:4-3}
 Offloading-decision making is the core of Phone2Cloud. It decides whether or not to offload computation of the application from smartphone to cloud, which is mainly used in the offloading decision engine. Fig. 2 shows the workflow of offloading-decision making algorithm. Table 1 lists the symbols and their meanings used in this paper.
\begin{figure}
\centering
\includegraphics [width=2.5in]{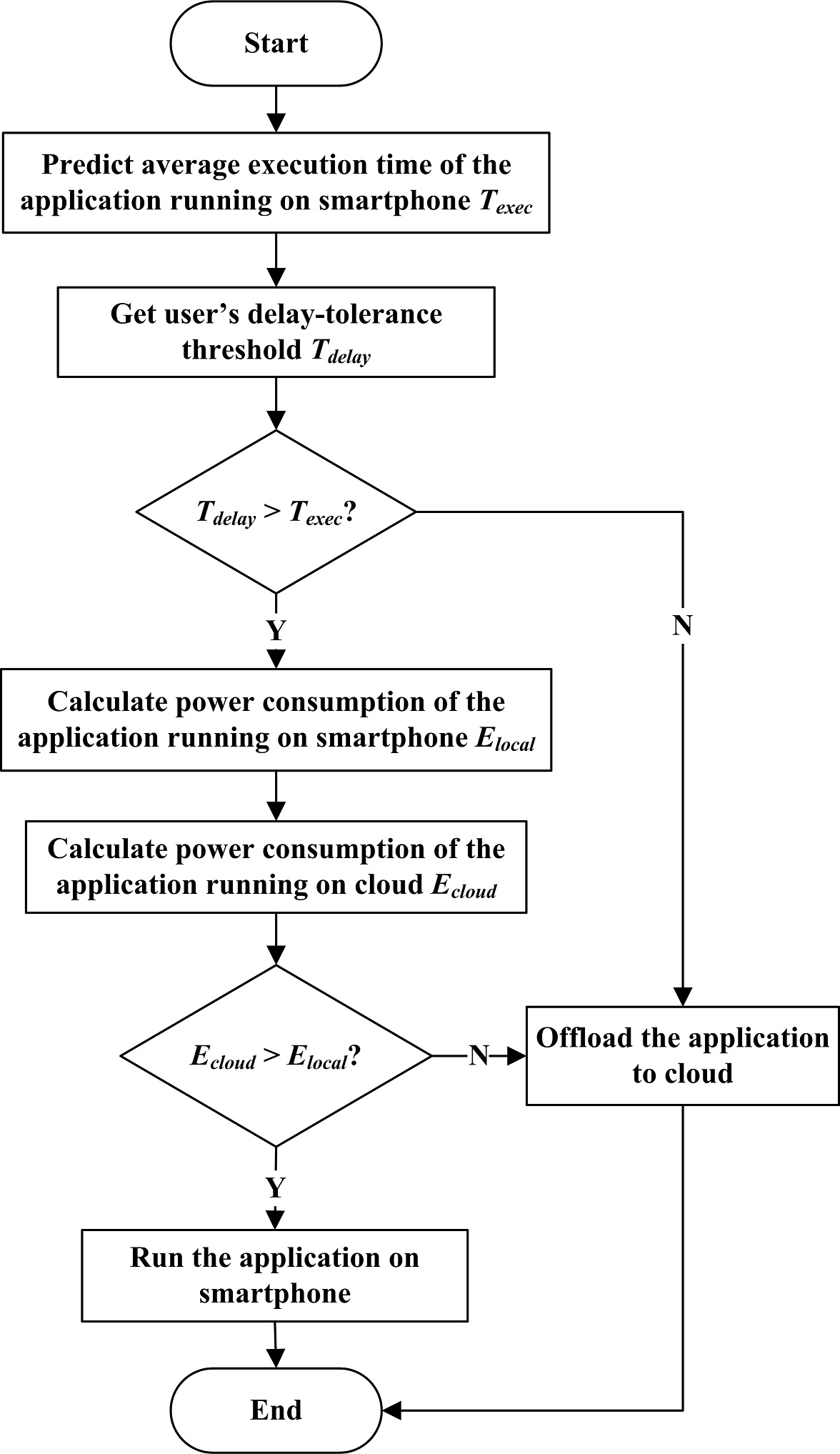}
\caption{Workflow of offloading-decision making algorithm}
\label{fig:1}
\end{figure}
\begin{table*}
\renewcommand{\arraystretch}{1.2}
\caption{List of notations}
\label{tab:1}
\begin{tabular*}{\textwidth}{@{\extracolsep{\fill}}l l l}
\hline
Variable & Description & Unit   \\
\hline
$C$	           & \multirow{1}{4in}{Computation complexity of an application}	 & I (instructions)                              \\
$M$            & \multirow{1}{4in}{Computation execution rate (speed) of smartphone} & I/s	                      \\
$S$	           & \multirow{1}{4in}{Computation execution rate (speed) of cloud}      & I/s                       \\
n	           & \multirow{1}{4in}{Ratio of computation execution rate on cloud and smartphone}             \\
$P_{exec}$	   & \multirow{1}{4in}{Active power of the processor on smartphone}	& W=J/s                          \\
$P_{idle}$	   & \multirow{1}{4in}{Idle power of smartphone (network interface \& processor included)}     & W=J/s \\
$P_{send}$	   & \multirow{1}{4in}{Power consumed to send data (network interface \& processor included)}  & W=J/s \\
$D_{send}$	   & \multirow{1}{4in}{Data size need to send to cloud} & B                                        \\
$B_{send}$	   & \multirow{1}{4in}{Sending bandwidth of the network}	& B/s                                      \\
$P_{receive}$  & \multirow{1}{4in}{Power consumed to receive data (network interface \& processor included)}   & W=J/s\\
$D_{receive}$  & \multirow{1}{4in}{Data size need to receive from cloud (execution results)}  & B              \\
$B_{receive}$  & \multirow{1}{4in}{Receiving bandwidth of the network}	                                  & B/s\\
$T_{exec}$	   & \multirow{1}{4in}{Execution time on smartphone}                                            & s\\
$T_{idle}$	   & \multirow{1}{4in}{Idle time on smartphone (execution time on cloud)}                       & s\\
$T_{send}$	   & \multirow{1}{4in}{Time for sending data}	                                                  & s\\
$T_{receive}$  & \multirow{1}{4in}{Time for receiving data}	                                              & s\\
$E_{local}$	   & \multirow{1}{4in}{Energy consumed by running application locally (on smartphone)}	  & J    \\
$E_{cloud}$    & \multirow{1}{4in}{Energy consumed by running application on cloud}	                      & J\\
$E_{tradeoff}$ & \multirow{1}{4in}{Trade-off energy consumed by computation offloading}	                  & J\\
\hline
\end{tabular*}
\end{table*}
Given an application, it needs five steps to make offloading decision:

\textbf{Step 1:} It gets the average execution time $T_{exec}$ of the application running on smartphone predicted by the execution time predictor.

\textbf{Step 2: }It gets user's delay-tolerance threshold $T_{delay}$ specified by user, and then compares $T_{exec}$ with $T_{delay}$. If user's delay-tolerance threshold is less than the average execution time of the application, then it decides to offload computation of the application from smartphone to cloud, expecting to buy some time for the application, and the algorithm ends. Otherwise, it goes to step 3.

\textbf{Step 3:} It calculates the power consumption of running the application on smartphone, called $E_{local}$, which will be described in Section 4.3.1.

\textbf{Step 4:} It calculates the power consumption of running the application on cloud, called $E_{cloud}$, and we will explain how to calculate this power consumption in Section 4.3.2.

\textbf{Step 5:} It compares $E_{local}$ with $E_{cloud}$. If $E_{cloud}$ is greater than $E_{local}$, then it decides to run the application on smartphone and then stops. Otherwise, it decides to offload computation of the application from smartphone to cloud, and the algorithm also ends.

In a nutshell, it basically compares two sets of variables ($T_{exec}$ and $T_{delay}$, $E_{local}$ and $E_{cloud}$) to make the offloading decision.
\subsubsection{Energy consumption on smartphone}
\label{sec:4-3-1}
In order to calculate the energy consumption consumed by running the application on smartphone, we need to know the active power of the processor on smartphone $P_{exec}$ and the average execution time $T_{exec}$ predicted by the execution time predictor, and then we use (7) to get power consumption $E_{local}$.
\begin{equation}
E_{local}=P_{exec}\cdot T_{exec}
\end{equation}
\subsubsection{Energy consumption on cloud}
\label{sec:4-3-2}
Calculating $E_{cloud}$ is more complicated than $E_{local}$. Before introducing how to calculate $E_{cloud}$, we make an assumption that the offloading part of the application is already on cloud when making an offloading decision.

As abovementioned, it needs three steps to finish computation offloading: sending the required input data, waiting the cloud completing execution of the offloaded computation and receiving execution results from cloud. Thus, $E_{cloud}$ includes three parts: the energy consumed by sending required input data on smartphone $E_{send}$, the energy of waiting execution results on smartphone $E_{idle}$ and the energy consumed by receiving execution results on smartphone $E_{receive}$.

$E_{send}$ is calculated by (8), and the time for sending data from smartphone to cloud $T_{send}$ is calculated by (9). Similarly, we calculate $E_{receive}$ by using (10) and (11).
\begin{equation}
E_{send}=P_{send}\cdot T_{send}
\end{equation}
\begin{equation}
T_{send}=D_{send}/B_{send}
\end{equation}
\begin{equation}
E_{receive}=P_{receive}\cdot T_{receive}
\end{equation}
\begin{equation}
T_{receive}=D_{receive}/B_{receive}
\end{equation}

It is easy to get (12), and we use it to represent execution time of the application on smartphone. Since we just offload whole or part of the application to cloud, we are easy to come to (13) and use it to present idle time on smartphone, i.e. execution time of the cloud-version application. Assume that \textit{n} represents the ratio of computation execution rate on cloud and smartphone, and then we get (14). Due to relatively large value of \textit{n}, we simply use the maximum value ${T_{idle}}_{max}$ to calculate the idle time on smartphone in our experiments. Therefore, we can use (14) and (15) to calculate $E_{idle}$.
\begin{equation}
T_{exec}=C/M
\end{equation}
\begin{equation}
T_{idle}\leq C/S
\end{equation}
\begin{equation}
T_{idle}\leq {T_{idle}}_{max}=T_{exec}/n
\end{equation}
\begin{equation}
E_{idle}=P_{idle}\cdot T_{idle}
\end{equation}

Thus, we can derive (16) to calculate $E_{cloud}$ from (8), (10) and (15). Furthermore, we get the trade-off energy consumption for computation offloading $E_{tradeoff}$ from (7) and (16), i.e. the difference of energy consumption of running the application on smartphone and cloud, as shown in (17).
\begin{eqnarray}
E_{cloud}&=&E_{send}+E_{idle}+E_{receive}\nonumber\\
         &=&P_{send}\cdot T_{send}+P_{idle}\cdot T_{idle}\\
         & &+P_{receive}\cdot T_{receive}\nonumber
\end{eqnarray}
\begin{eqnarray}
E_{tradeoff}&=&E_{local}-E_{cloud}\nonumber\\
            &=&P_{exec}\cdot T_{exec}-P_{send}\cdot T_{send}\\
            & &-P_{idle}\cdot T_{idle}-P_{receive}\cdot T_{receive}\nonumber
\end{eqnarray}

We transform (17) to (18) and (19). Then we make $E_{tradeoff}$ equal to zero and get a constant, called break-even transmission energy $E_{0}'$, as shown in (20). Once the application and the status of smartphone are specified, the value of $E_{0}'$ is constant.
\begin{equation}
E_{tradeoff}=P_{exec}\cdot T_{exec}-P_{idle}\cdot T_{idle}-E'
\end{equation}
\begin{equation}
E'=P_{send}\cdot T_{send}+P_{receive}\cdot T_{receive}
\end{equation}
\begin{equation}
E_{0}'=P_{exec}\cdot T_{exec}-P_{idle}\cdot T_{idle}
\end{equation}

In summary, we use (14) and (20) to calculate $E_{0}'$, use (9), (11) and (19) to calculate the variable $E'$ and then compare $E_{0}'$ with $E'$. If $E_{0}'$ is greater than $E'$, Phone2Cloud offloads computation of the application from smartphone to cloud. Otherwise, it runs the application on smartphone.
\section{Evaluation and analysis}
\label{sec:5}
In this section, we conduct two different sets of experiments: application experiments and scenario experiments. Then we analyze the energy consumption and execution time of the applications in our experiments under four different factors, including input size, bandwidth, CPU workload and delay-tolerance threshold.
\subsection{Experiment setup}
\label{sec:5-1}
The application and scenario experiments are based on an environment depicted in Fig. 3. We use ZTE V880 smartphone in our experiments. It uses Android operating system in version 2.2, integrates with Wi-Fi interface and is capable of EDGE data connectivity. It has a Qualcomm MSM7227-1 CPU with 600MHz frequency, a 256MB memory and a battery capacity of 1250mAh at 3.7 volts. The mobile client part of Phone2Cloud, including resource monitor, bandwidth monitor, execution time predictor, offloading decision engine, local execution manager, and offloading proxy, will be run on it. As we can see from Fig. 3, we use Wi-Fi network to connect cloud in our experiments.
\begin{figure}
\centering
\includegraphics [width=2.5in]{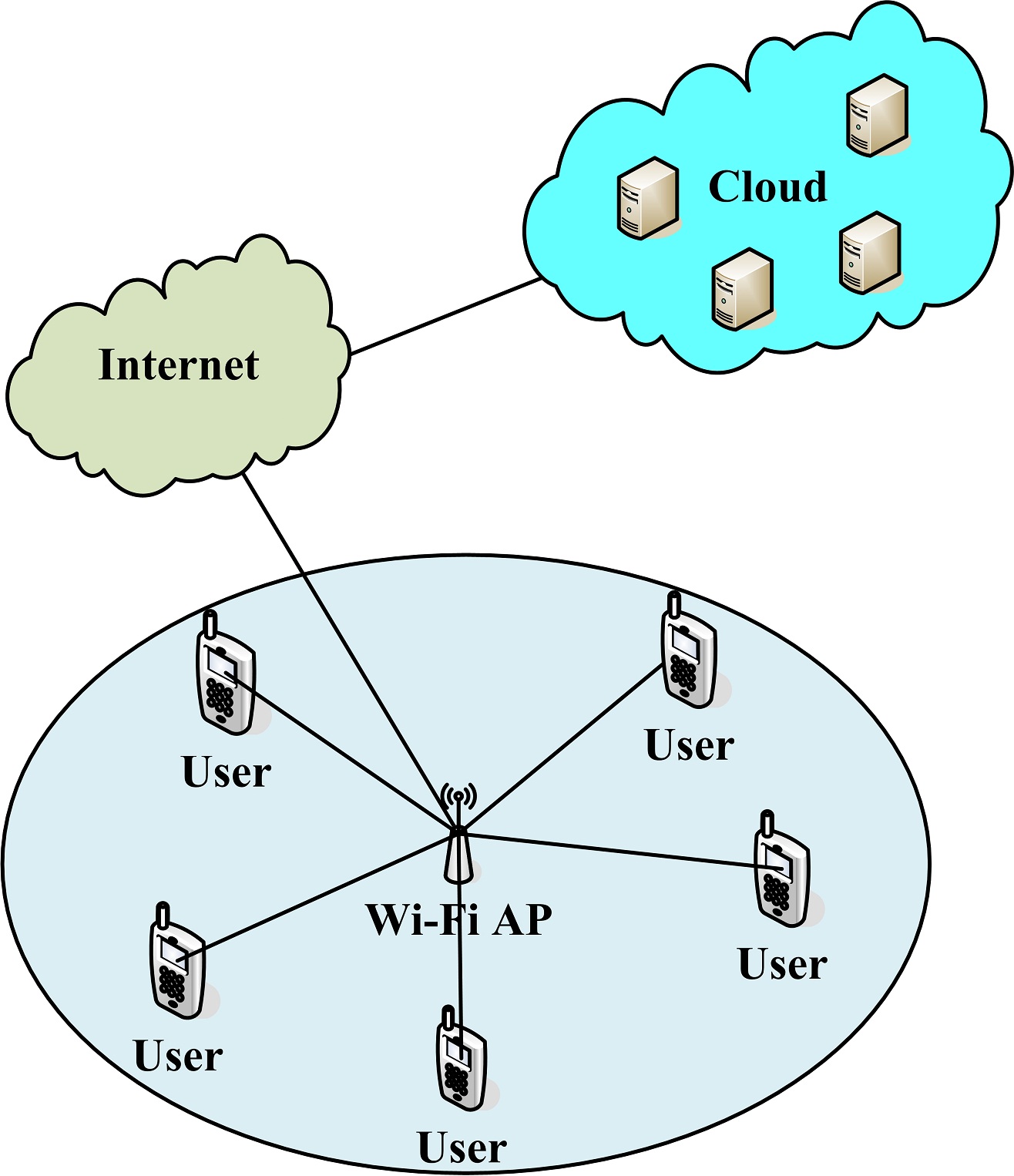}
\caption{Environment used in application experiments and scenario experiments}
\label{fig:1}
\end{figure}
As to the cloud, we use a desktop computer running Hadoop in version 0.20.2 in Linux operating system (Ubuntu 10.04) to serve as a cloud. It has a Pentium dual-core CPU with 2.60GHz frequency, a 2GB memory, a 320GB hard disk and a 100Mbps network interface. Besides, we run the cloud part of Phone2Cloud, i.e. remote execution manager, on the desktop computer.

We use three applications, as shown in Table 2, to do application experiments and analyze the energy consumption and execution time of three applications from the experimental data.
\begin{table}
\renewcommand{\arraystretch}{2}
\caption{Applications used in experiments}
\label{tab:2}
\begin{tabular*}{3.3in}{l lp{0.3in} lp{2.6in}}
\hline
No. & Application & Description  \\
\hline
\multirow{1}*{1} & \multirow{1}*{sort}        & \multirow{1}{2.15in}{Sort a given set of integer array elements by using Quick sort}	 \\
\multirow{2}*{2} & \multirow{2}*{path finder} & \multirow{2}{2.15in}{Given a map and a source location (node), finds the shortest path tree with the source location as root by using Bellman-Ford algorithm}\\
~~&~~&~~\\
\multirow{1}*{3} & \multirow{1}*{word count}  & \multirow{1}{2.15in}{Count the number of words from a block of text}     \\
\hline
\end{tabular*}
\end{table}
The scenario experiments are based on a scenario that many students are using smartphones to learn an image process course via Internet. In our scenario experiments, we assume that they simply use an application running on smartphones, called \textit{face finder}, to find the number of faces in a picture. Moreover, we should run the above four applications under different input sizes and CPU workloads as much as possible before experiments, so that we can get a large history log \textit{L} used for execution time prediction described in Section 3.2.

We also utilize two experimental tools to record our experimental data and change CPU workload with minimal intrusion to our system respectively. One is called \textit{PowerUsage} \cite{39}, which is used to measure power consumption of smartphones. It uses battery interfaces provided by Google APIs to record power consumption of smartphones. The other one is called \textit{CPUChanger}, and we develop it to change CPU workload on demand.
\subsection{Analysis of application experiments}
\label{sec:5-2}
We evaluate both energy consumption and execution time of three applications in Table 2 with respect to four factors. We will examine how these factors will affect energy consumption and execution time of the applications.

Table 3 describes the four factors' values in application experiments for evaluations of both energy consumption and execution time. For a specified factor, we evaluate its influence on both energy consumption and execution time of three applications under different ranges, and other three factors are set to their default values.
\begin{table*}
\renewcommand{\arraystretch}{2}
\caption{Parameters in application experiments}
\label{tab:3}
\begin{tabular*}{\textwidth}{@{\extracolsep{\fill}}l l l l l l l l l}
\hline
\multirow{2}*{Application}	&\multicolumn{2}{l}{Input size}&\multicolumn{2}{l}{CPU workload}& \multicolumn{2}{l}{Delay-tolerance threshold}   &\multicolumn{2}{l}{Bandwidth}\\
~~ & \multirow{1}{0.3in}{Range (KB)}   & \multirow{1}{0.3in}{Default (B)}   & \multirow{1}{0.3in}{Range (\%)}
   & \multirow{1}{0.3in}{Default (\%)} & \multirow{1}{0.3in}{Range (ms)}    & \multirow{1}{0.3in}{Default (ms)}
   & \multirow{1}{0.3in}{Range (KB/s)} & \multirow{1}{0.3in}{Default (KB/s)}  \\
\hline
Sort        & 0$\scriptsize{\sim}$4000 & 40000   & \multirow{3}*{0$\scriptsize{\sim}$80}     & \multirow{3}*{51.36}
            & 40$\scriptsize{\sim}$140 & \multirow{3}*{infinite}
            & \multirow{3}*{0$\scriptsize{\sim}$800}    & \multirow{3}*{731.50}\\
Path finder & 0$\scriptsize{\sim}$250  & 101,111 & ~~ & ~~& 4000$\scriptsize{\sim}$22000 \\		
Word count  & 0$\scriptsize{\sim}$2000 & 524,337 & ~~ & ~~& 2000$\scriptsize{\sim}$4000  \\			
\hline
\end{tabular*}
\end{table*}
\subsubsection{Energy consumption}
\label{sec:5-2-1}
In this section, we show the connections between energy consumption of three applications and four factors including input size, bandwidth, CPU workload, and delay-tolerance threshold. As a matter of fact, most of computation of these applications can be offloaded to the cloud, so the data needed to be sent to the cloud is the input data of these applications, and it is reasonable that we use the maximum value ${T_{idle}}_{max}$ as the idle time on smartphone in Section 4.3.2. As we mentioned before, our system is not a fully automatic system, and it needs us to manually modify applications, so that their offloading parts can be run on the cloud and they can receive execution results returned from the cloud. Thus, there are three corresponding cloud-version applications.

\textbf{Input size}

Fig. 4 shows the results of the power consumption of three applications with different input sizes, based on the values in Table 3. In the figure, offloading (the green line) represents the power consumption of different three applications running in Phone2Cloud, which has the same indications in Figs. 5 - 7.
\begin{figure*}
\centering
\includegraphics [width=6.6in]{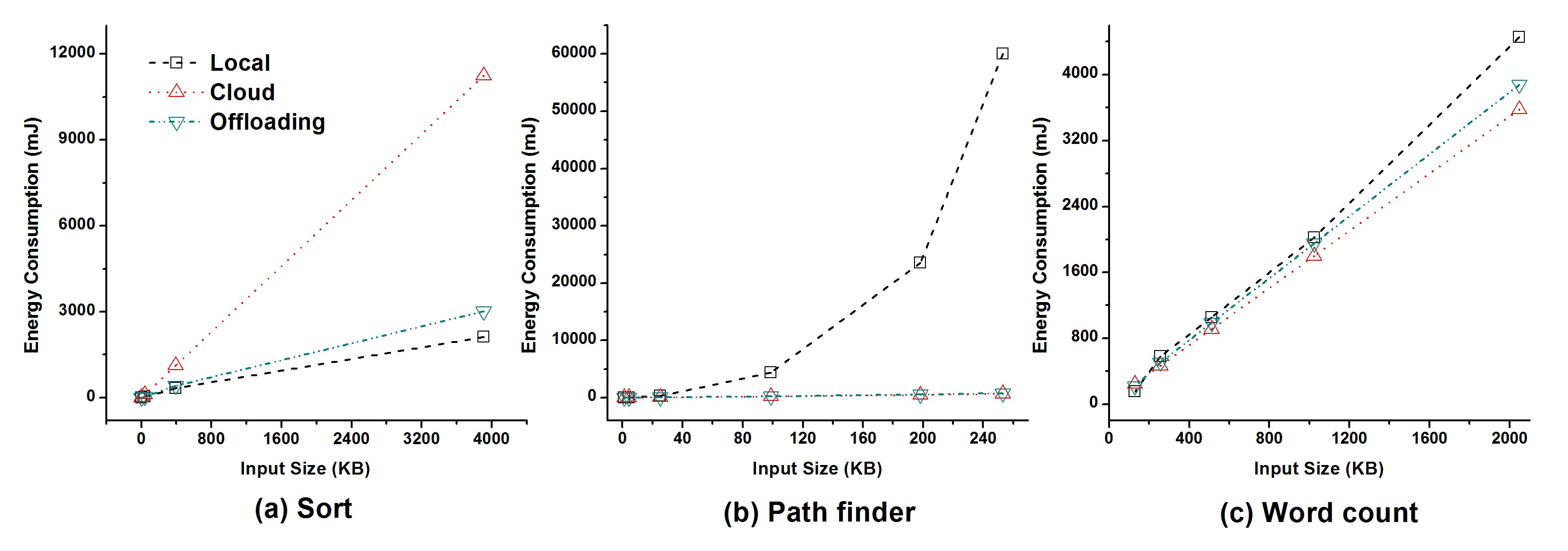}
\caption{Energy consumption of three applications under different input sizes}
\label{fig:4}
\end{figure*}
As we can see from Fig. 4(a), the energy consumed by \textit{sort} running on cloud is much more than that on smartphone. The reason is that energy consumed by sorting on smartphone is much less than that by data transmission, including sending input data and receiving results. We also see that \textit{sort} should be always directly run on smartphone. Due to the overhead of Phone2Cloud itself, such as the power consumption of making the offloading decision, Phone2Cloud consumes a little more power than running \textit{sort} on smartphone, which can be seen in Fig. 4(a). Moreover, overhead of Phone2Cloud can be seen among all figures.

On the contrary, the power consumption of running \textit{path finder} on smartphone is much more than running it on cloud, as we can see from Fig. 4(b). Because of the high complexity of finding a path in a map and relatively small data transmission, the energy consumed by searching paths in a map on smartphone is much more than transmitting data, and this leads to the above result. We also observe that \textit{path finder} should be always offloaded to cloud. Moreover, we can see that our system save much energy for \textit{path finder} under these circumstances.

As to \textit{word count}, the result is interesting, as shown in Fig. 4(c). When input size is smaller than 256KB, the application running on cloud costs more energy than running locally. After that, the power consumption on smartphone is over that on cloud. There is a reason for this result. When the input size is smaller than 256KB, transferring data costs more power than counting words. The more the input size is, the more power is used to counting words. We also see that \textit{word count} should be run on smartphone when input size is smaller than 256KB and should be offloaded to cloud when input size is larger than 256KB. Therefore, we can save energy for \textit{word count} when input size is larger than 256KB, and it increases as input grows.

\textbf{Bandwidth}
\begin{figure*}
\centering
\includegraphics [width=6.6in]{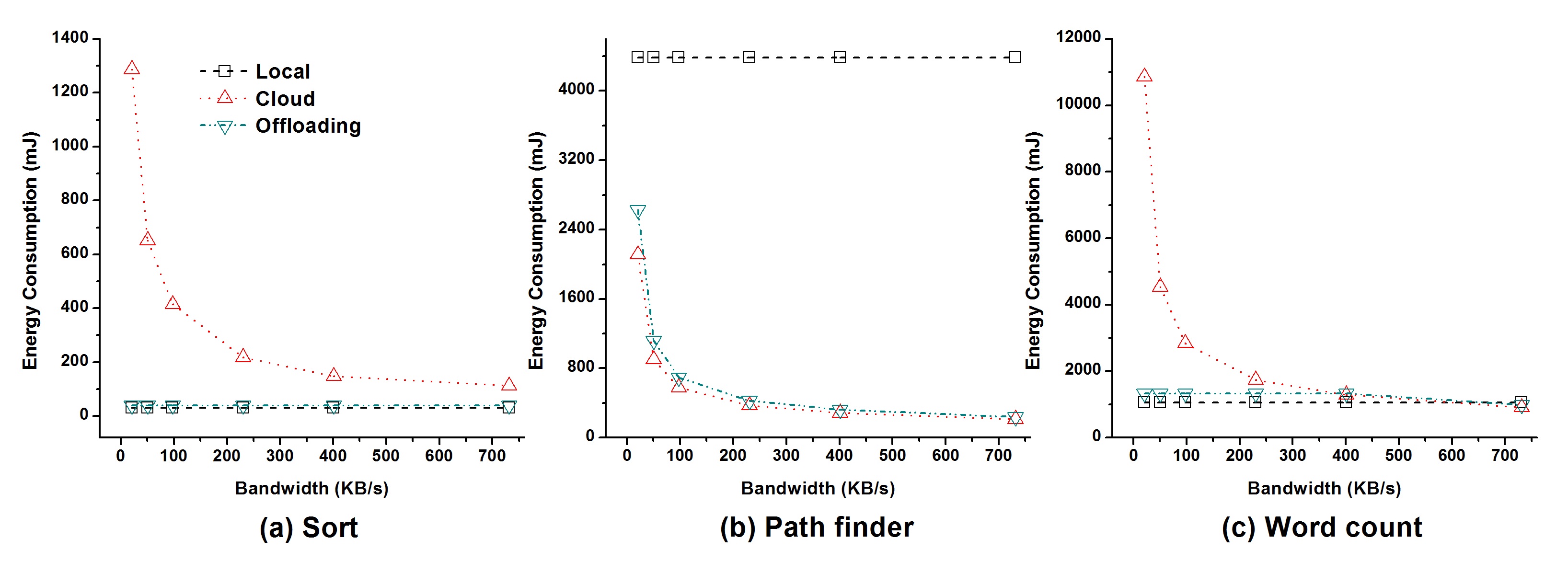}
\caption{Energy consumption of three applications under different bandwidths}
\label{fig:5}
\end{figure*}

Fig. 5 describes how bandwidth affects three applications' energy consumption. For \textit{sort} in Fig. 5(a), the value of power consumption on cloud is always higher than that on smartphone, that is to say, \textit{sort}  should be always run on smartphone. The explanation for this is energy consumed by sorting computation on smartphone is less than that consumed by data transmission. However, energy consumption of \textit{sort} running on cloud and smartphone are getting closer and closer with bandwidth increasing. Furthermore, the power consumed on smartphone keeps the same and we can also see this in Figs. 5(b) and 5(c).

We can see from Fig. 5(b) that power consumption of \textit{path finder} on smartphone is much more than that on cloud. The reason is the same as that of Fig. 4(b). Additionally, the power consumption on cloud is decreasing with bandwidth increasing, and we see the same in Fig. 5(c). So we should always offload \textit{path finder} to cloud under such circumstances, and then we can save much energy for users.

The results in Fig. 5(c) are similar to those in Fig. 5(a), but there is a difference. That is \textit{word count} consumes more energy on cloud compared with smartphone when bandwidth is below 600KB/s, while the power consumption on cloud is less than smartphone when bandwidth gets higher. The reason is that the power consumption of data transmission is getting less and less. Furthermore, we can reach that \textit{word count} should be run on smartphone when bandwidth is below 600KB/s and offloaded to cloud when bandwidth is higher. Therefore, we can save energy for \textit{word count} when the bandwidth is greater than 600KB/s, and the benefit enlarges as bandwidth increases.

\textbf{CPU workload}
\begin{figure*}
\centering
\includegraphics [width=6.6in]{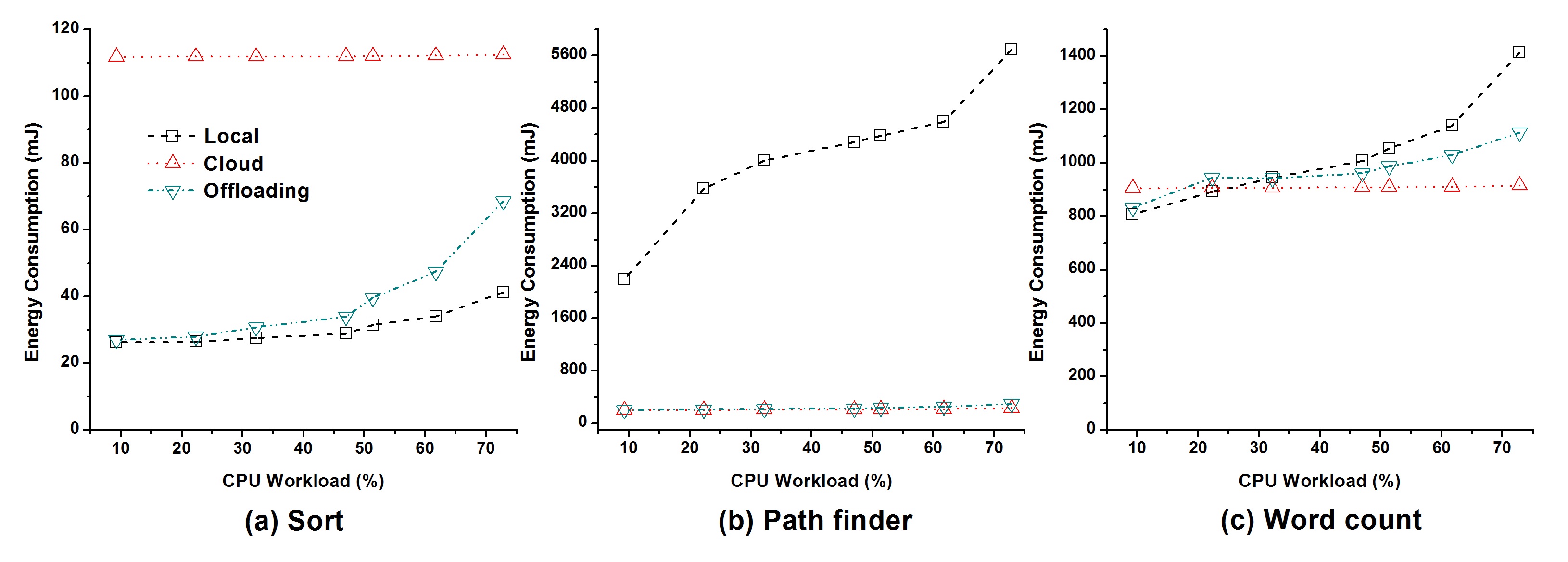}
\caption{Energy consumption of three applications under different CPU workloads}
\label{fig:6}
\end{figure*}

The power consumption of three applications under different CPU workloads is clearly shown in Fig. 6. In Fig. 6(a), we can see that energy consumed by \textit{sort} running on cloud is always more than running on smartphone, meaning the application should be always run on smartphone under these situations. Moreover, the power consumption of \textit{sort} running on smartphone is getting more and more with CPU workload growing. As CPU workload increases, the energy consumption of smartphone for waiting execution results from cloud grows. However, the growth is too small to be seen in Fig. 6(a). Furthermore, we can also see such situations in Figs. 6(b) and 6(c). Fig. 6(b) witnesses that power consumption of \textit{path finder} running on smartphone is always much more than that on cloud, and it increases as CPU workload grows. Therefore, we should always offload \textit{path finder} to cloud under these situations to save energy. However, Fig. 6(c) shows a very different situation compared with Fig. 6(b) when it comes to \textit{word count}. Its power consumption on smartphone is less than that on cloud when CPU workload is below 25\%, and the opposite occurs when CPU workload is above 25\%. The reason is that energy consumption of \textit{word count} running on smartphone grows with CPU workload increasing and energy consumption of running on cloud almost does not change. Therefore, \textit{word count} should be offloaded to cloud when CPU workload is greater than 25\%, and we can save energy for users.

\textbf{Delay-tolerance threshold}
\begin{figure*}
\centering
\includegraphics [width=6.6in]{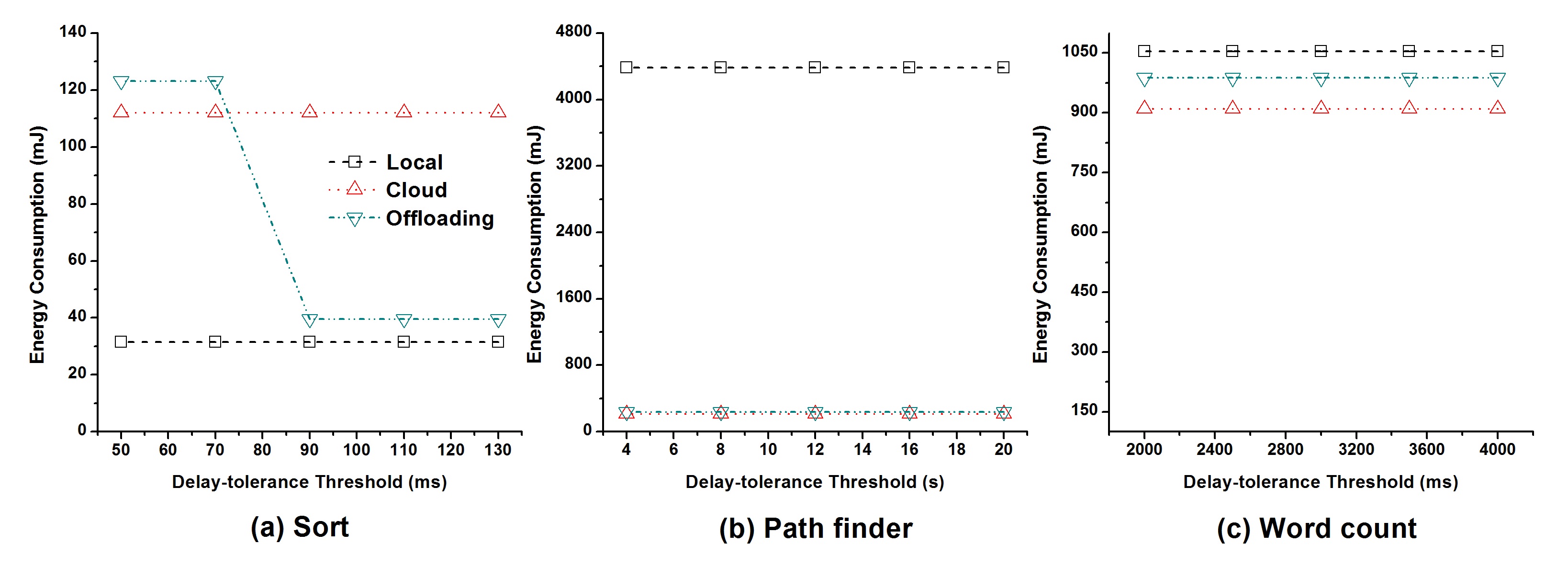}
\caption{Energy consumption of three applications under different delay-tolerance thresholds}
\label{fig:7}
\end{figure*}

Fig. 7 illustrates the energy consumption of three applications under different delay-tolerance thresholds. We see that all of three applications experience the same phenomenon: the energy consumption of different applications running on cloud does not change with delay-tolerance threshold growing as well as running on smartphone. For this phenomenon, we can explain that delay-tolerance threshold does not affect three applications' power consumption on cloud and smartphone. However, the power consumption of \textit{sort} running in our system is changing as delay-tolerance grows in Fig. 7(a). According to Fig. 4(a), the power consumption of \textit{sort} running in our system should be close to running on smartphone all the time. However, it is close to the power consumption of running on cloud when delay-tolerance threshold is smaller than 70 milliseconds, which means the execution time of \textit{sort} running on smartphone is between 70 milliseconds and 90 milliseconds. Therefore, we should set delay-tolerance threshold to be greater than 90 milliseconds for \textit{sort}. As shown in Figs. 7(b) and 7(c), \textit{path finder} and \textit{word count} should be offloaded to cloud all the time; however, we can not know their execution time of running on smartphone.

\subsubsection{Execution time}
\label{sec:5-2-2}
We describe the relationship between the execution time of three applications and four factors mentioned before in this section. The values of each factor for different applications are given in Table 3, and the results are shown in Figs. 8 - 10. In these figures, offloading (the green line) represents the execution time of different applications running in our Phone2Cloud.

\textbf{Input size}
\begin{figure*}
\centering
\includegraphics [width=6.6in]{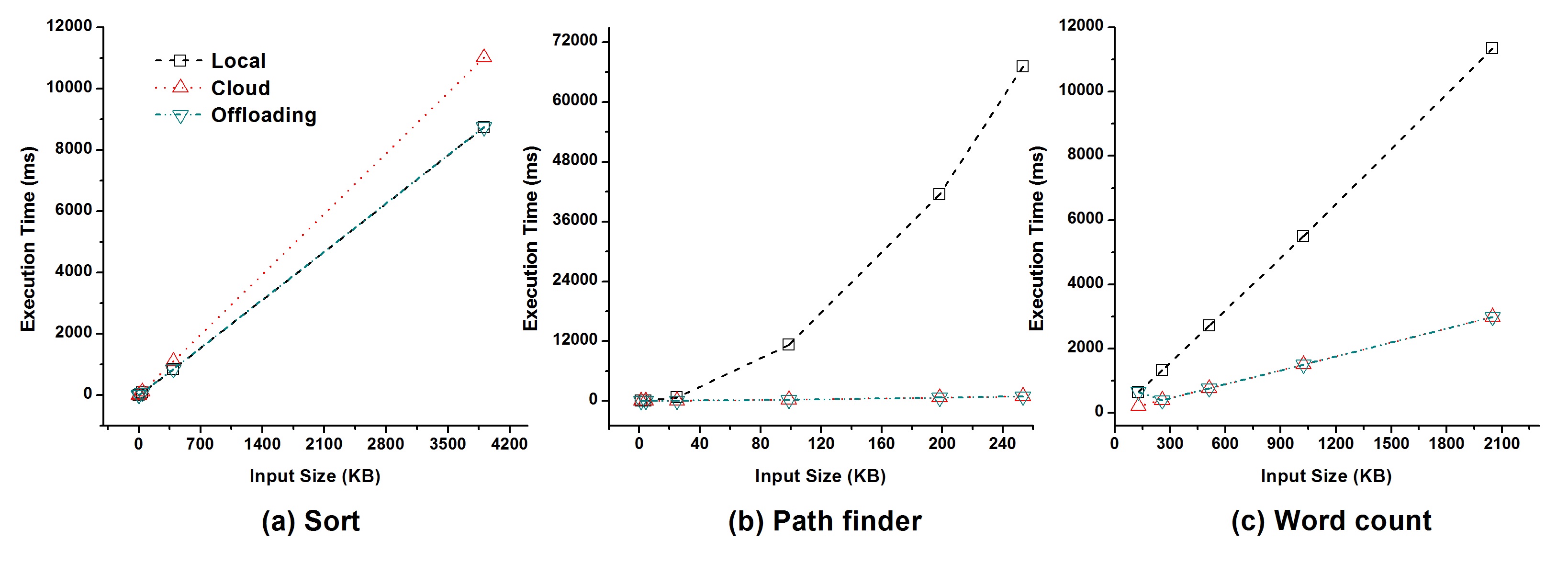}
\caption{Execution time of the three applications under different input sizes}
\label{fig:8}
\end{figure*}

Fig. 8(a) shows the execution time of \textit{sort} under different input sizes. The value on cloud is more than that on smartphone, and the gap between them expands as input grows. The reason lies in that time cost by data transmission between cloud and smartphone is more than that cost by sorting on smartphone. Therefore, \textit{sort} should be run on smartphone on this condition. Except for this, we see that the offloading one keeps the same as the local one. In this case, we see our system can make a wise offloading decision.

For \textit{path finder}, the execution time on smartphone overweighs that on cloud, and the difference becomes more obvious as input increases, shown in Fig. 8(b). As a result, we should run \textit{path finder} on cloud. Besides, the execution time on cloud increases a little, but the change is too tiny to be seen in the figure. It can be seen that our system can make excellent decisions for \textit{path finder}.

The wisdom of Phone2Cloud can also be proved by Fig. 8(c). Due to the fact that execution time of \textit{word count} on cloud is less than that on smartphone when input size is smaller than 256KB, we should run the application on cloud. While, the fact is that our Phone2Cloud runs \textit{word count} on smartphone, and it is a wise decision. As we already know that user's delay-tolerance threshold is infinite, so the first task of our Phone2Cloud is to reduce the application's energy consumption. Furthermore, the power cost on cloud is more than that on smartphone when input size is below 256KB in Fig. 4(c). Therefore, Phone2Cloud decides to run \textit{word count} on smartphone. When input size is greater than 256KB, our Phone2Cloud still makes a wise decision, thus user's experience is improved.

\textbf{Bandwidth}
\begin{figure*}
\centering
\includegraphics [width=6.6in]{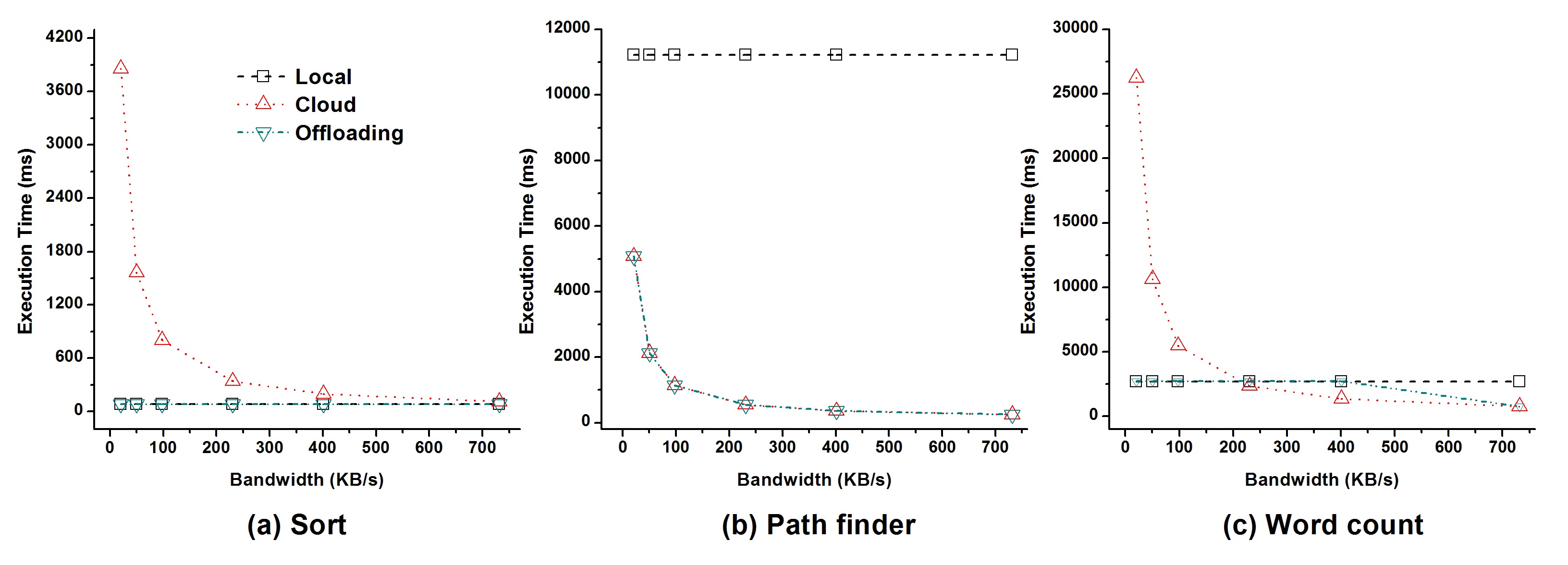}
\caption{Execution time of three applications under different bandwidths}
\label{fig:9}
\end{figure*}

Fig. 9 shows the relationship between execution time of three applications and bandwidth. The running on cloud always consumes more time than running locally, shown in Fig. 9(a). The reason is that sorting on smartphone costs less time than data transmission. Moreover, the gap between running on cloud and locally becomes smaller as bandwidth grows. We can see from Fig. 9(a) that the execution time on Phone2Cloud almost keeps the same as that on smartphone, regardless of the overhead of Phone2Cloud. This shows our system always makes the right decision.

Fig. 9(b) shows a different result comparing with Fig. 9(a). \textit{Path finder} spends more time on smartphone than on cloud. This is because the time of data transmission is less than finding a path in a map on smartphone, and it decreases as bandwidth increases. Therefore our system chooses to offload \textit{path finder} to cloud, and in this situation, we can save time for users.

For \textit{word count}, the execution time on cloud is more than that on smartphone when bandwidth is less than 230KB/s, while the opposite occasion occurs when bandwidth is greater. But that does not mean we should offload the application to cloud  even though it could save time for users. Considering the energy consumption of \textit{word count} on cloud and smartphone, our system offloads \textit{word count} to cloud only when the bandwidth exceeds 600KB/s, and it makes the right decision again.

\textbf{CPU workload}

Fig. 10 shows the execution time of \textit{sort}, \textit{path finder} and \textit{word count} under different CPU workloads. A notable feature shared by the subfigures is that the execution time of three applications running on cloud does not change with CPU workload growing.
\begin{figure*}
\centering
\includegraphics [width=6.6in]{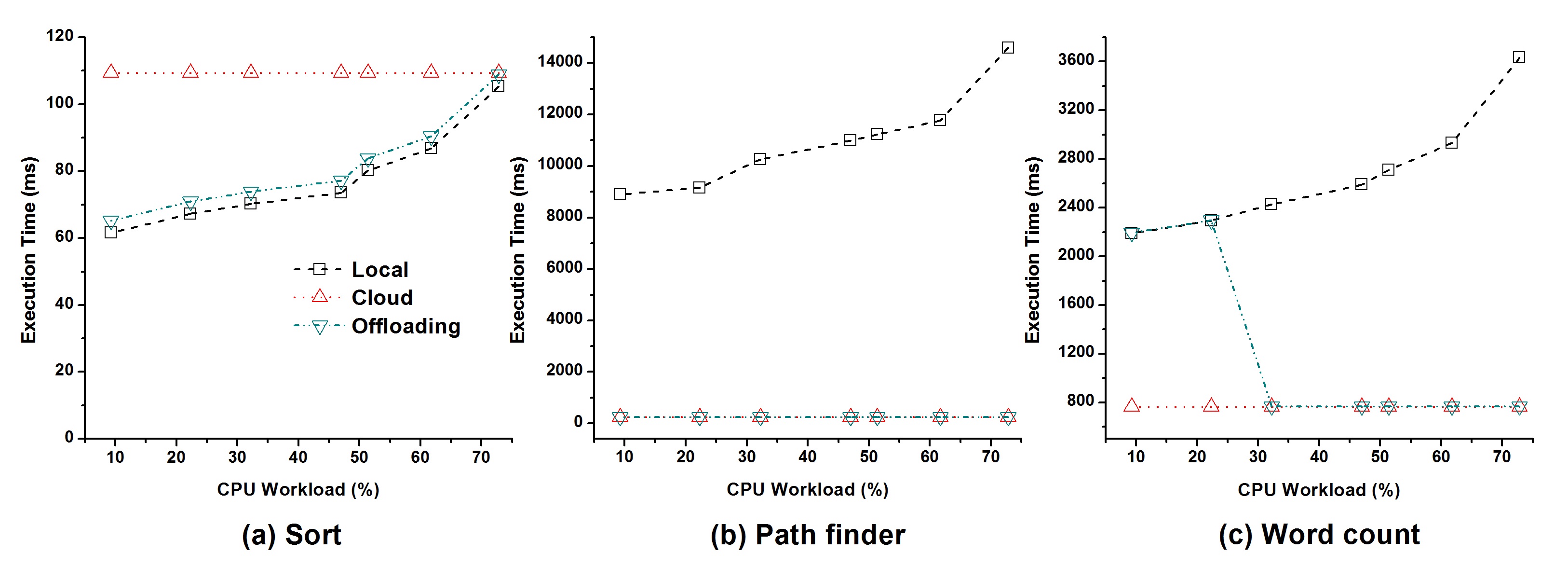}
\caption{Execution time of three applications under different CPU workloads}
\label{fig:10}
\end{figure*}
For \textit{sort} in Fig. 10(a), the time consumed on cloud is consistently higher than that on smartphone. Furthermore, the gap between running on smartphone and cloud reduces when CPU workload increases. However, the execution time of \textit{path finder} on cloud keeps much less than on smartphone and this is quite different from \textit{sort} (see Fig. 10(b)). Additionally, the execution time on smartphone grows as CPU workload increases, so we should always offload \textit{path finder} to cloud under these situations to save time for users. As to \textit{word count}, the results in Fig. 10(c) correspond to the power consumption in Fig. 6(c). Due to infinite of delay-tolerance threshold, users pay much attention on energy consumption, so Phone2Cloud runs \textit{word count} locally when CPU workload is lower than 25\% from Fig. 6(c). While CPU workload is greater than 25\%, it is offloaded to cloud to save energy and time, as shown in Fig. 10(c).

\textbf{Delay-tolerance threshold}
\begin{figure*}
\centering
\includegraphics [width=6.6in]{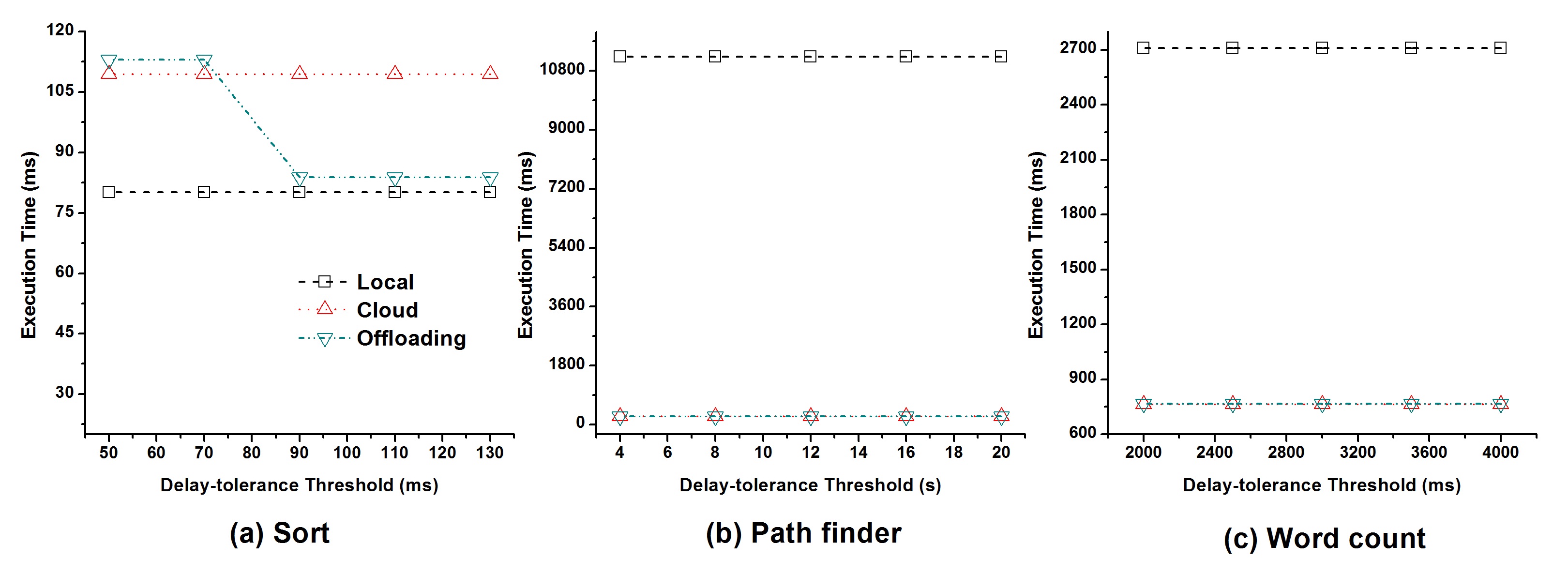}
\caption{Execution time of three applications under different delay-tolerance thresholds}
\label{fig:11}
\end{figure*}

Fig. 11 describes the relation between execution time and delay-tolerance threshold. We can see that the trend this figure shows is similar to Fig. 7. Because there is no connection between delay-tolerance threshold and execution time on smartphone and cloud, the execution time of each application on both cloud and smartphone does not change with delay-tolerance threshold growing. Besides, the execution time of \textit{sort} in Phone2Cloud changes as delay-tolerance threshold increases, similar to the energy consumption in Fig. 7(a). Thus, we get the conclusion that the execution time of \textit{sort} on smartphone is about 80 milliseconds as shown in Fig. 11(a). Therefore, delay-tolerance threshold should be larger than 90 milliseconds, to be precise, 80 milliseconds. In this way, time can be saved. Based on what Fig. 4(b) shows, we can understand that the energy consumption of \textit{path finder} on smartphone is more than on cloud. Therefore, no matter how much delay-tolerance threshold is, our Phone2Cloud always offloads this application to cloud. Specifically, the local execution time of \textit{path finder} is around 11 seconds, while both cloud and offloading ones are less than 1 second, thus we achieve the goal of saving time for users. Fig. 11(c) shows very similar results to Fig. 11(b). \textit{Word count} is always offloaded to cloud in regardless of delay-tolerance threshold. Also, time can be saved by Phone2Cloud for \textit{word count} as shown in Fig. 11(c).

\subsection{Analysis of scenario experiments}
\label{sec:5-3}
This section examines the energy consumption and execution time of \textit{face finder} in scenario experiments under the aforementioned factors. The connections between energy consumption and factors will be discussed in Section 5.3.1, and the results on execution time will be presented  in Section 5.3.2.

Table 4 shows each factor's range and default value in the scenario experiments for evaluations of both energy consumption and execution time. When we evaluate a specified factor's influence on both energy consumption and execution time of face finder under different ranges, the other three factors are set to their default values.
\begin{table}
\renewcommand{\arraystretch}{1.5}
\caption{Parameters in scenario experiments}
\label{tab:4}
\begin{tabular}{l l l l l}
\hline
\multirow{2}{0.15in}{Factor}  & \multirow{2}{*}{Input size} & \multirow{2}{0.2in}{CPU workload} &
\multirow{2}{0.4in}{Delay-tolerance threshold} & \multirow{2}{0.2in}{Bandwidth}  \\
 ~~ & ~~  &  ~~  &  ~~  & ~~ \\
\hline
Range   & 0$\scriptsize{\sim}$1200 KB & 20$\scriptsize{\sim}$80 \%  & 10$\scriptsize{\sim}$50 s     & 0$\scriptsize{\sim}$800 KB/s  \\
Default & 559,064 B  & 51.36 \%     & infinite                  & 20.27 KB/s  \\
\hline
\end{tabular}
\end{table}
\subsubsection{Energy consumption}
\label{sec:5-3-1}
Before discussing the results shown in Fig. 12, we need to manually modify \textit{face finder} to be able to receive results from cloud, and there is also a corresponding cloud-version \textit{face finder} on cloud. Since most of computation of \textit{face finder} can be offloaded to the cloud, the required data of the cloud-version \textit{face finder} is the input data of \textit{face finder}, and it is also reasonable that we use the maximum value ${T_{idle}}_{max}$ to calculate the idle time on smartphone.
\begin{figure}
\centering
\includegraphics [width=3.3in]{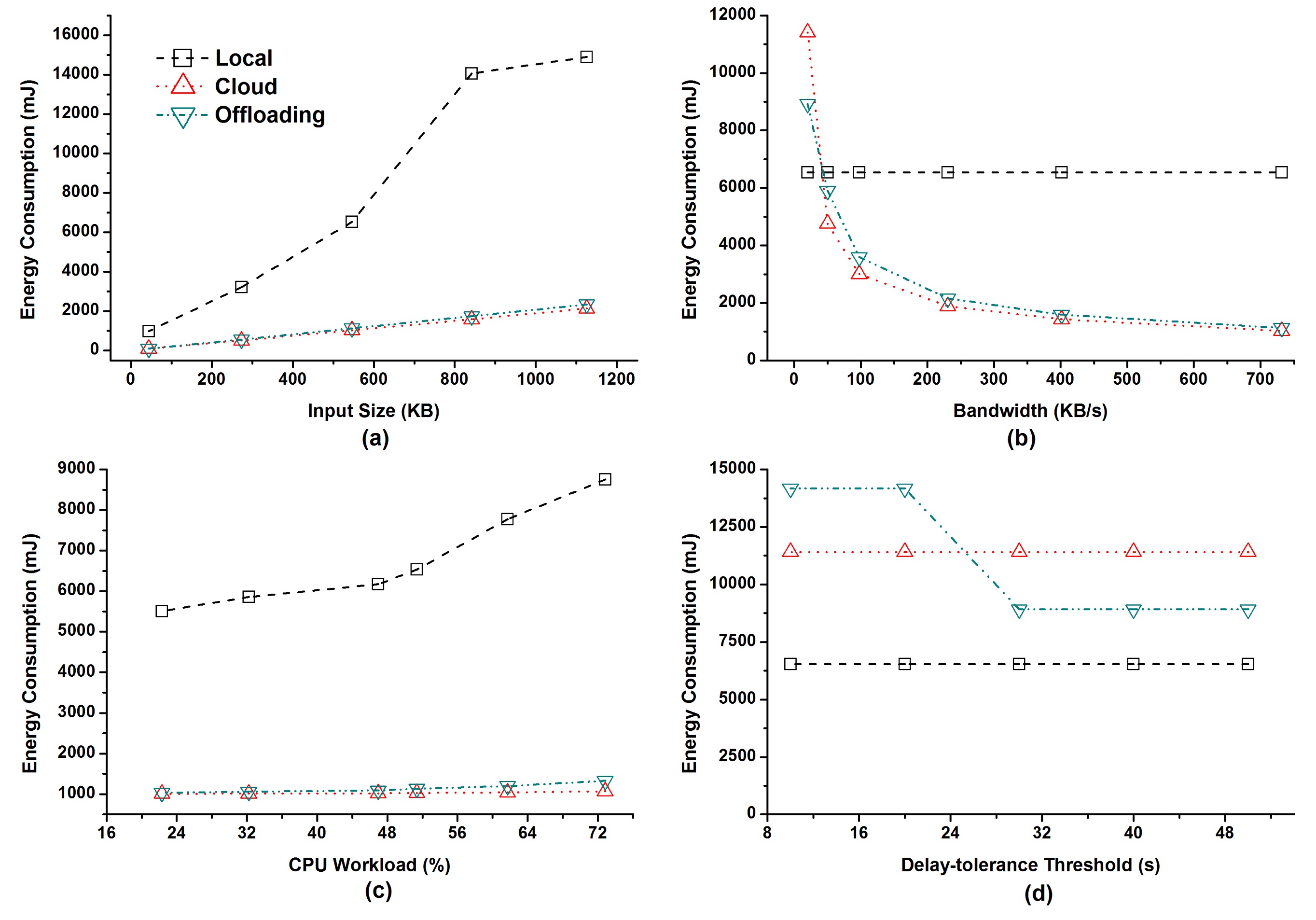}
\caption{Energy consumption of face finder under four factors}
\label{fig:12}
\end{figure}

Fig. 12(a) shows that \textit{face finder} costs more energy on smartphone than on cloud and the difference between them gets larger as input grows. This is because data transmission costs less energy than running the application locally. Furthermore, the local energy consumption grows faster than that on cloud. Therefore, \textit{face finder} should be offloaded to cloud.

From Fig. 12(b), we can see that \textit{face finder} cost less power on smartphone than on cloud when bandwidth is lower than 50KB/s and the result changes to the opposite after bandwidth gets higher. The former is because the energy consumed by data transmission is very enormous when bandwidth is low, due to the poor power efficiency of Wi-Fi. Similar to what Fig. 9(a) shows, the power consumption on smartphone has no relation to bandwidth, and the power consumed on cloud decreases with bandwidth growing. As a consequence, we should offload \textit{face finder} to cloud when bandwidth is greater than 50KB/s for energy saving.

As to the effect of CPU workload, Fig. 12(c) shows that energy consumption of \textit{face finder} on cloud is always much less than running locally, and the gap between them grows as CPU workload increases. So \textit{face finder} should be offloaded to cloud, which is what our system does.

For the delay-tolerance threshold, it nearly affects nothing on \textit{face finder} on both smartphone and cloud as shown in Fig. 12(d). According to energy consumption of \textit{face finder}, Phone2Cloud should run \textit{face finder} on smartphone when the threshold is smaller than 30 seconds. However, it offloads the application to cloud, meaning the execution time on smartphone can not meet user's threshold. So the local execution time is between 20 seconds and 30 seconds. Therefore, to save power, delay-tolerance threshold should be greater than 30 seconds.
\subsubsection{Execution time}
\label{sec:5-3-2}
Fig. 13 describes the execution time of \textit{face finder} under the four factors.
\begin{figure}
\centering
\includegraphics [width=3.3in]{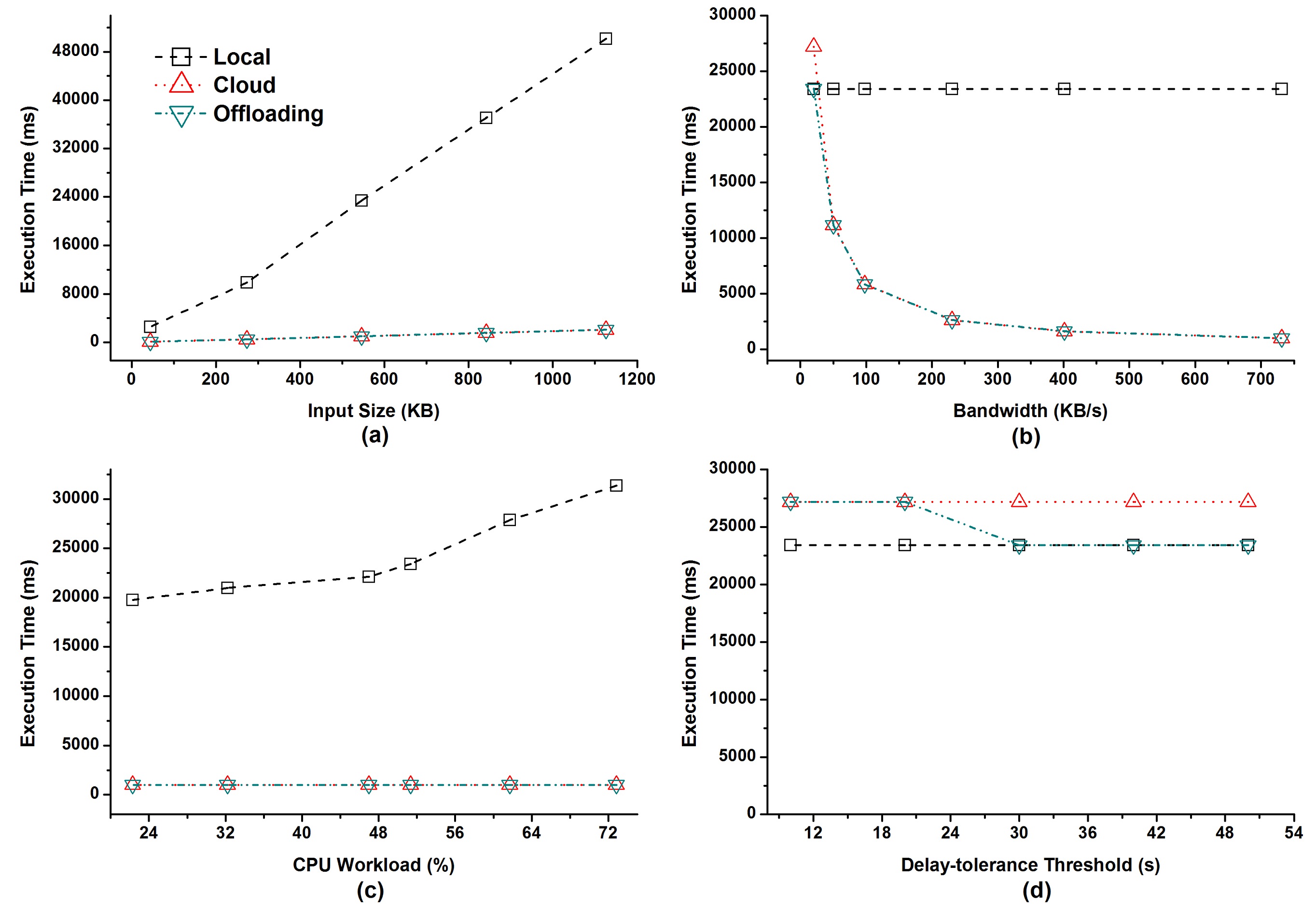}
\caption{Execution time of face finder under four factors}
\label{fig:13}
\end{figure}
The subfigures in Fig. 13 illustrate the connection between execution time and the four factors respectively. To be specific, the execution time on smartphone is not only more than, but also grows faster than that on cloud, as shown in Fig. 13(a). Therefore, Phone2Cloud offloads \textit{face finder} to cloud all the time, meaning it makes a wise offloading decision.

As clearly shown in Fig. 13(b), we can see the execution time experiences almost the same trend with the energy consumption in Fig. 12(b), caused by the inefficient data transmission under low bandwidth. Furthermore, the execution time on cloud decreases with bandwidth ascending and there is no relationship between local execution time and bandwidth. So \textit{face finder} should be offloaded to cloud when bandwidth is greater than 50KB/s, as we stated in Fig. 12(b).

In Fig. 13(c), we also see the similar trend shown in Fig. 12(c), with the only difference that the execution time on cloud is a straight line, meaning CPU workload influences nothing on execution time on cloud. Then in this case, our Phone2Cloud makes an excellent offloading decision, that is, it offloads \textit{face finder} to cloud, and saves time for users.

As regard to delay-tolerance threshold, Fig. 13(d) validates our thought again, which is the threshold does not correlate to the execution time on both cloud and smartphone. The execution time in Phone2Cloud changes with delay-tolerance threshold increasing. Ignoring the threshold, \textit{face finder} should be run on smartphone according to its energy consumption in Fig. 12(b). However, Phone2Cloud offloads it to cloud when delay-tolerance threshold is smaller than 25 seconds. This means the execution time of \textit{face finder} on smartphone is between 20 seconds and 30 seconds, as what we got from Fig. 12(d). Actually, it is about 23 seconds shown in Fig. 13(d). Therefore, we should set delay-tolerance threshold to be greater than 23 seconds to save time for users.

\section{Conclusion}
\label{sec:6}
This paper has presented an energy-efficient mobile cloud computing system called Phone2Cloud, which takes advantage of the computation offloading paradigm. It is able to save energy and improve applications' performance and users' experience of smartphones. Two sets of experiments are conducted and the results demonstrate that our system is of great effectiveness. For the sake of simplicity, we simply use a naive approach to predict application's execution time and only take CPU workload and input size into account. There are many alternative methods and factors we can take into consideration, and we will compare these methods with our own approach in the future work. Moreover, as we mentioned before our system is a semi-automatic offloading system. Further studies are demanded to make our system fully automatic.

\section*{Acknowledgment}
This work was partially supported by the Natural Science Foundation of China under Grant No. 60903153 and No. 61203165, Liaoning Provincial Natural Science Foundation of China under Grant No. 201202032, and the Fundamental Research Funds for the Central Universities (DUT12JR10).

\end{document}